\documentclass{article}

\usepackage{microtype}
\usepackage{graphicx}
\usepackage{subcaption}
\usepackage{booktabs} 

\usepackage{xurl}
\usepackage{hyperref}

\usepackage[preprint]{icml2026}

\usepackage{amsmath}
\usepackage{amssymb}
\usepackage{mathtools}
\usepackage{amsthm}
\usepackage{times}
\usepackage{latexsym}
\usepackage{xcolor}
\usepackage[dvipsnames]{xcolor}
\usepackage[skins,breakable]{tcolorbox}
\usepackage{longtable}
\usepackage{cuted}
\usepackage{enumitem}
\usepackage{verbatim}
\usepackage{amsmath}
\usepackage[normalem]{ulem}
\tcbuselibrary{breakable}
\usepackage{booktabs}
\usepackage{array}
\usepackage{todonotes}
\definecolor{PromptBlue}{RGB}{85, 150, 210}    
\definecolor{PromptGreen}{RGB}{110, 170, 120}  
\definecolor{PromptOrange}{RGB}{215, 150, 90}  
\definecolor{PromptPurple}{RGB}{150, 120, 185}  
\definecolor{PromptTeal}{RGB}{90, 160, 165}     

\usepackage[capitalize,noabbrev]{cleveref}

\theoremstyle{plain}

\theoremstyle{definition}

\theoremstyle{remark}

\icmltitlerunning{The Biosecurity Blind Spot}

\begin{document}

\twocolumn[
    \icmltitle{The Biosecurity Blind Spot: Systematic Dual-use Detection in Open Science Infrastructure}



  \icmlsetsymbol{equal}{*}

  \begin{icmlauthorlist}
    \icmlauthor{Vasudha Sharma}{comp}
    \icmlauthor{Chakresh Kumar Singh}{yyy}
    \icmlauthor{Jayesh Choudhari}{yyy}
    \icmlauthor{Dharmit Nakrani}{yyy}
  \end{icmlauthorlist}

  \icmlaffiliation{yyy}{Independent}
  \icmlaffiliation{comp}{\href{https://lophilabs.ai/}{LophiLabs}}

  \icmlcorrespondingauthor{Vasudha Sharma}{vasudha.sharma@lophilabs.com}
  \icmlcorrespondingauthor{Chakresh Kumar Singh}{chakresh.kr.singh@gmail.com}
  \icmlcorrespondingauthor{Jayesh Choudhari}{jayesh.choudhari17@gmail.com}

  \icmlkeywords{Machine Learning, ICML}

  \vskip 0.3in
]



\printAffiliationsAndNotice{}  

\begin{abstract}
AI is transforming life sciences research at unprecedented speed, accelerating discovery across protein structure prediction, genome modeling, and drug development (\citet{jumper2021highly, mak2024artificial}).
Yet this rapid advancement, coupled with the open science movement, introduces significant dual-use research concerns that have received limited empirical scrutiny.
Here we present the first systematic analysis of dual-use research of concern (DURC) content on open preprint servers.
We screened ~52,000 bioRxiv preprints (2024–2025) using a hybrid pipeline of lexical filtering and large language model (LLM) evaluation, scoring metadata across nine DURC, three PEPP, and five governance categories aligned with U.S. and Australia Group oversight frameworks.
Our analysis reveals that dual-use-adjacent knowledge is routinely present in openly accessible titles and abstracts, often exceeding established risk thresholds even in studies with legitimate public health objectives.
While this mapping captures surface-level information diffusion, it does not measure operational capability, downstream misuse potential, or the substantial technical and biosafety barriers that constrain harmful application.
We argue that institutional review processes, funding requirements, and preprint platform policies must evolve to incorporate proactive, metadata-level monitoring without compromising scientific transparency.
Ultimately, harmonizing controlled-access mechanisms for high-risk methodologies with open summaries of scientific contributions offers a pragmatic framework for governing AI-accelerated biology at scale.
\href{https://anonymous.4open.science/r/biorisk_blind_spot-877D/}{{\color{blue}{[Code Repo]}}}
\end{abstract}

\begin{figure}[t]
  \centering
  \includegraphics[width=0.80\linewidth]{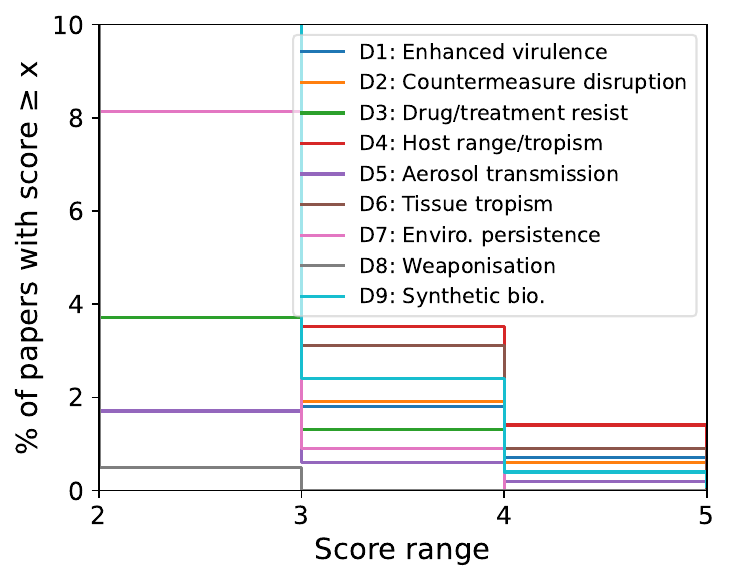}
  \caption{
  Complementary cumulative distribution function (CCDF) of mean scores across nine DURC criteria (D1-D9) for 1{,}000 bioRxiv preprints, showing the proportion of papers scoring at or above each threshold $t \in [2,5]$. 
  The most prevalent DURC indicators at $t=3$ are D4 (Altered Host Range: 3.5\%) and D6 (Altered Tissue Tropism: 3.1\%), reflecting technical detail on host-range expansion and tissue-specific infection mechanisms. 
  The steep decline beyond $t=3$ indicates that high-severity dual-use content is concentrated in a small subset of preprints. 
  Distributions for PEPP (P1--P3) and governance (G1--G5) criteria are provided in Appendix Figure~\ref{app:fig:ccdf_full}.}
  \label{fig:ccdf}
\end{figure}
\section{Introduction}

The convergence of artificial intelligence and the life sciences has catalyzed a structural shift in biological research, dramatically compressing discovery timelines across protein structure prediction, genome-scale modeling, and therapeutic design (\citet{jumper2021highly, mak2024artificial, baek2021rosettafold, dauparas2022proteinmpnn, lin2023language, corso2023diffdock, chen2022interpretable, olsen2022observed, shuai2021generative, bagal2022molgpt, ahmad2022chemberta, madani2020large, ferruz2022protgpt2}). 
This acceleration is amplified by the open science movement, particularly the widespread adoption of preprint servers like bioRxiv, which prioritize rapid, unrestricted dissemination. 
While this culture has democratized access and fostered global collaboration, it simultaneously lowers the barrier to acquiring dual-use knowledge: information that can be leveraged for public benefit or repurposed for harmful applications.

Dual-use research of concern (DURC) has long been recognized as a critical challenge in the life sciences, yet governance remains fragmented. 
Fewer than 25 countries maintain formal DURC oversight frameworks, and international coordination lags behind technological capability (\citet{national2004biotechnology, who2022global, casebookgreene2023biorisk}). 
The integration of generative AI into research workflows further complicates this landscape: AI systems now routinely assist in literature synthesis, experimental design, and the identification of pathogenic sequences, raising concerns about the unintentional proliferation of sensitive knowledge through open channels (\citet{brundage2018malicious}). 
Although measures to flag DURC material in publications exist, they remain weak and largely ineffective for preprints. The Materials Design Analysis Reporting (MDAR) framework, piloted across journals from 2017–2019, required authors to disclose DURC oversight in a free-text box but lacked structure to capture how biorisks were assessed and placed the reporting burden solely on authors at publication (Greene et al., 2023; Macleod et al., 2021). For preprint servers like bioRxiv, these limitations are amplified, as researchers can bypass stricter journal review entirely by routing work through platforms with looser or absent dual use oversight (Greene et al., 2023). Compounding these limitations, journal-level DURC policies, including bioRxiv's,  have not been meaningfully updated since the rise of generative AI, leaving a widening governance gap as biological AI models and LLMs produce dual-use outputs (e.g., novel protein designs, synthetic sequences) that fall outside the agent- and pathogen-based scope of existing DURC and PEPP frameworks (Bloomfield et al., 2024; Wang et al., 2025; Pannu et al., 2025).
Despite growing policy discourse, empirical analyses of how dual-use indicators manifest in publicly accessible literature remain scarce.
Existing studies rely heavily on theoretical risk assessments or post-hoc incident reviews, leaving a critical gap in our understanding of the actual prevalence and distribution of potentially sensitive content in real-time research outputs.

For instance, a recent preprint by \citet{gracias2026entry} investigates bat coronavirus spillover, a legitimate public health goal—yet discloses technical detail on viral immune suppression and host-range expansion. 
Applied post-hoc, our framework flagged this title-abstract for \textit{Altered Host Range (D4)}, \textit{Altered Tissue Tropism (D6)}, and \textit{Pandemic Potential (P1)}, with average LLM scores $>3/5$ across three independent runs.
BioRxiv's submission screening prioritizes speed over the in-depth risk assessment typical of formal journals; our results illustrate that algorithmic screening can surface dual-use-adjacent content even when human review is time-constrained. 
This highlights a structural gap between legitimate inquiry and unintended knowledge diffusion, not authorial negligence.

In this (ongoing) work, we present the first systematic analysis of dual-use research indicators across 50,000 bioRxiv preprints published between 2024 and 2025. We employ a hybrid screening pipeline that combines initial keyword filtering with LLM evaluation to classify papers across three oversight dimensions: established DURC categories (n=9), PEPP indicators (n=3), and governance-relevant markers (n=5). 
Our objective is not to equate academic publication with malicious intent, but to empirically map the availability of technical knowledge that, when combined with declining barriers to synthesis, automation, and commercial lab access, could lower the threshold for misuse. 
We also confront the methodological and ethical limitations inherent in this approach: string-based filtering inevitably misses semantically relevant but lexically divergent content, LLM-as-judge frameworks carry known calibration biases, and the presence of DURC-adjacent information does not, by itself, constitute an imminent threat. 
Nevertheless, the findings underscore a structural tension between open science norms and biosecurity imperatives, particularly in academic settings where commercial oversight is minimal and incentive structures prioritize novelty over risk mitigation. 
We argue that institutional review processes, funding agency requirements, and preprint platform policies must evolve to incorporate proactive, lifecycle-spanning DURC monitoring without compromising the transparency that drives scientific progress.

\section{Methods \& results}
Our dataset comprises 52,713 de-duplicated English-language bioRxiv preprints (2024–2025) with extracted metadata (DOI, title, abstract, authors, date); all analyses operated exclusively on titles and abstracts without full-text retrieval. This data set is processed via a two stage (i) screening and, (ii) flagging pipeline. 

\noindent\textbf{Stage~1: Screening through lexical filtering.} 
We applied deterministic string matching against agent names, technical protocols, and policy terms (see Appendix~\ref{app:keywords}) from the 2024 U.S. DURC/PEPP frameworks \citep{ostp2024durcpepp} and the 2025 Australia Group Common Control List \citep{australiagroup_handbook_volii}. 
This high-recall filter retained 2{,}361 of 52{,}713 preprints (4.5\%) for the evaluation of Stage~2; the remainder were excluded as low-relevance. Lexical false positives (e.g., ``anthrax'' in historical epidemiology) are expected to be filtered during automated scoring in stage 2, although residual misclassifications may persist. 
Of these 2{,}361 we chose 1000 most recent papers as our test sample for the purpose of this ongoing work. 

\noindent\textbf{Stage~2: LLM classification \& flagging outcomes.} 
A large language model (GPT-4.1) evaluated each title and abstract pair across 17 criteria: nine DURC indicators (D1-D9), three PEPP markers (P1-P3), and five governance features (G1-G5). Using structured prompts and category-specific rubrics (see Appendix~\ref{app:prompts}), the model assigned integer scores (0-5) per category based on presence, specificity, and technical depth of dual-use relevant content. Each assessment was repeated three times with permuted category orderings to mitigate anchoring \citep{choudhari2025prompt}; final scores represent the mean across runs. 
A preprint is classified as \emph{flagged} if the mean score on any criterion reaches $\geq 3$.

\begin{figure}[h]
  \centering
  \includegraphics[width=0.9\linewidth]{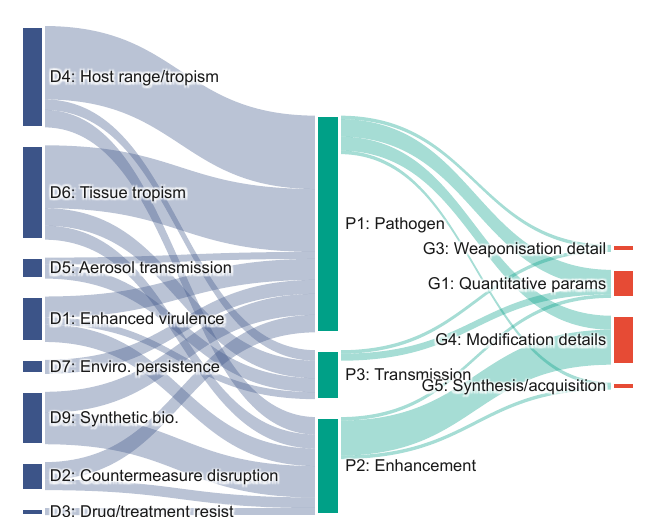}
  \caption{\textbf{Co-occurrence of dual-use indicators.} Sankey diagram of papers scoring $\geq 3$ across DURC (D1--D9), PEPP (P1--P3), and governance (G1--G5) criteria. Of 142 PEPP-flagged papers, 56 also trigger DURC and 17 add governance concerns; 13 satisfy all three. This pattern indicates a subset of literature concurrently encodes pathogen risk, technical enhancement detail, and sensitive operational parameters.}
  \label{fig:sankey}
\end{figure}

Of the 1{,}000 preprints evaluated, 232 (23.2\%) were flagged on at least one criterion: 114 (11.4\%) on DURC, 142 (14.2\%) on PEPP, and 53 (5.3\%) on governance. Among the flagged papers, 56 scored $\geq 3$ on both DURC and PEPP dimensions, and 13 satisfied all three simultaneously. Score distributions for DURC criteria are shown in Figure~\ref{fig:ccdf}; PEPP and governance distributions appear in Appendix Figure~\ref{app:fig:ccdf_full}. Co-occurrence patterns across all dimensions are shown in Figure~\ref{fig:sankey}.

\begin{figure}[h]
  \centering
  \includegraphics[width=\linewidth]{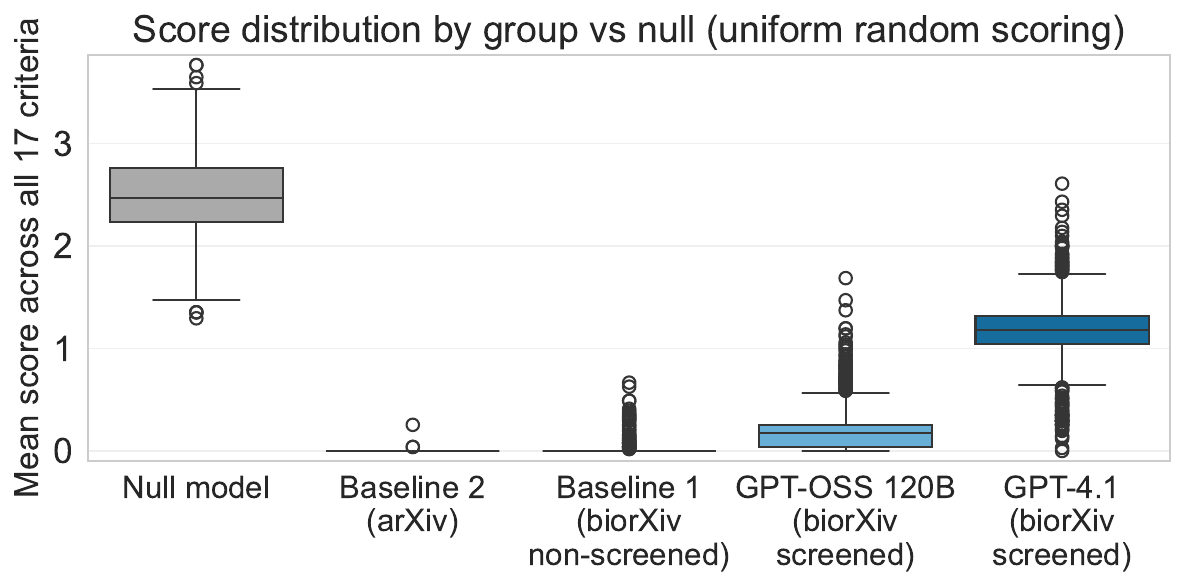}
  \caption{Score distributions for each group against a discrete uniform
  random-scoring null ($\mathcal{U}\{0,1,2,3,4,5\}$, theoretical mean~2.5).
  Each box shows the distribution of per-paper mean scores across all
  17 criteria. Groups ordered left to right by increasing mean.
  All groups differ significantly from the null ($p < 0.001$, two-sided
  permutation test, 5{,}000 draws, seed~42). Box plot with five groups: null model, Baseline 2 (arXiv),
  Baseline 1 (non-flagged bioRxiv), GPT-OSS 120B, and GPT-4.1 flagged
  cohort. The null box is centred near 2.5. All four observed groups
  sit well below the null; arXiv baseline is closest to zero.}
  \label{fig:permtest}
\end{figure}

\noindent\textbf{Negative-control calibration}
To assess the correctness of categorization by the LLMs and effective callibration, we set-up negative control datasets for our pipelines. We constructed two such datasets namely:
\begin{itemize}
    \item bioRxiv [non-screened]: 500 random papers from the cohort that were not screened
    in stage 1. This data set serves dual purpose. As these papers did not have the agents from our screening list from stage I, we expect them to not have any information that should be flagged. Therefore, they should perform low on the overall scoring and control for the residual misclassification. A low score on this data should mean that our pipeline works as expected. 
    \item arXiv : To further strengthen our control, we also fetch 500 completely random pre prints from arXiv. Most of the arXiv preprints should fall out of the scope for this study, and thus we expect the least scoring on this dataset. 
    However, we also agree that  there could be residual information even in arXiv preprints, e.g.,  papers from computer science and quantitative biology.
\end{itemize}
The structure of the metadata fetched was kept same as the papers in our data-set, and we use two LLMs for our scoring. One is an open weight modes gpt-oss:120b and other gpt-4.1. Permutation tests (two-sided, null~ $= \mathcal{U}\{0{-}5\}$, theoretical
mean ~2.5) confirm that all four groups diverge significantly from uninformed
random scoring ($p < 0.001$). The ordering arXiv ($\bar{x}=0.001$) $<$ non-screened bioRxiv ($\bar{x}=0.020$) $<$ flagged bioRxiv on GPT-oss:120b ($\bar{x}=0.19$) $\ll$ flagged bioRxiv on GPT-4.1 ($\bar{x}=1.162$) demonstrates bidirectional
discrimination. The model exhibits suppression of scores for non-relevant content
(both baselines lie far below the 2.5 random baseline), while assigning
substantially higher scores to keyword-flagged biomedical pre-prints.
The gap between Baseline~1 (same biology domain) and Baseline~2 (non-biology)
further confirms that the model responds to domain relevance, not merely to document length or vocabulary richness.
These results are visualized in Figure~\ref{fig:permtest}.
All four observed groups fall significantly \emph{below} the null,
confirming the model suppresses scores for non-relevant content rather
than defaulting to mid-range values. The ordering Baseline~2 (arXiv)
$<$ Baseline~1 (non-screened bioRxiv) $<$ flagged cohort demonstrates
discrimination at both the domain level (biology vs.\ non-biology) and
within biology (general vs.\ dual-use-relevant research).

\subsection{Inter-run reliability}
To assess how consistently our judge models [gpt-oss, GPT-4.1] score the same paper across independent runs, we computed Krippendorff's $\alpha$ (ordinal scale) per criterion. $\alpha$ compares observed disagreement between runs to the disagreement expected under chance. $\alpha = 1$ can be interpreted as perfect agreement across runs, $\alpha = 0$ reflects agreement by chance,  and $\alpha < 0$ indicates systematic disagreement. Krippendorff recommends $\alpha \geq 0.80$ for reliable conclusions and $\alpha \geq 0.67$ as the minimum for tentative, exploratory findings~\cite{krippendorff2019}.

\begin{figure}[h]
  \centering
  \includegraphics[width=\linewidth]{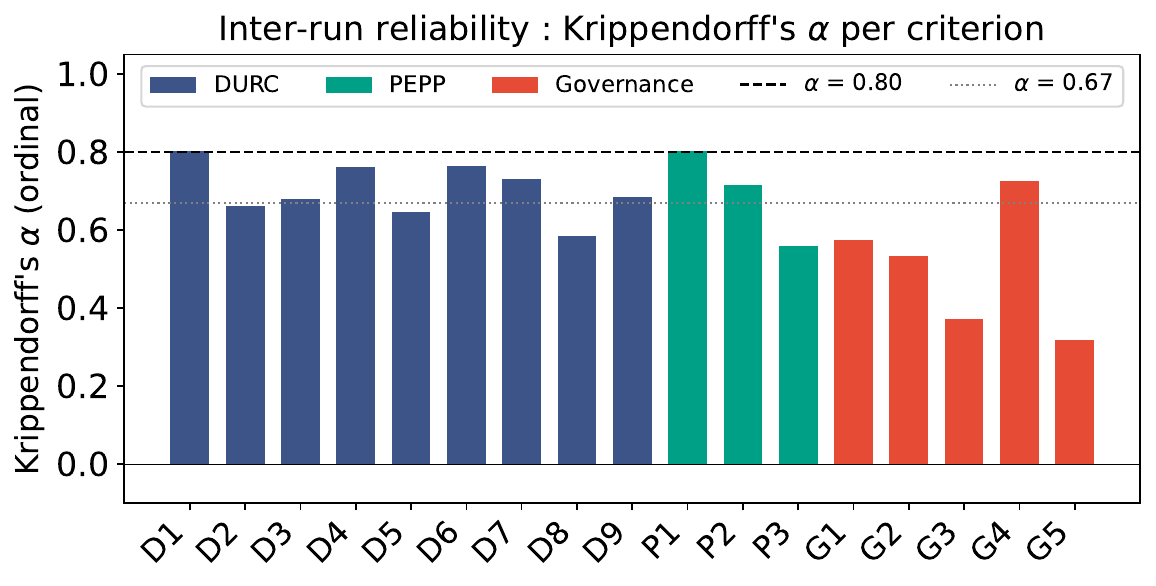}
  \caption{Krippendorff's $\alpha$ (ordinal) per criterion for GPT-4.1
  (dark bars) and GPT-OSS~120B (light bars).
  Dashed line: $\alpha = 0.80$ (reliable conclusions);
  dotted line: $\alpha = 0.67$ (acceptable for exploratory work).
  Governance criteria G1--G5 are operationally defined for this study
  and have not been externally validated. Grouped bar chart with 17 criterion groups on the x-axis
  and Krippendorff alpha on the y-axis. Two bars per criterion compare
  GPT-4.1 (dark) and GPT-OSS 120B (light). Horizontal dashed line at
  0.80 and dotted line at 0.67 mark reliability thresholds. Most PEPP
  and key DURC criteria exceed 0.67; governance criteria G3 and G5
  are the lowest, below 0.40.}
  \label{fig:kripp}
\end{figure}

Across our 17 criteria, $\alpha$ ranged from 0.33 to 0.80 (Figure~\ref{fig:kripp} and Table~\ref{tab:kripp}).
Eight criteria met the $\alpha \geq 0.67$ exploratory threshold, including
key DURC indicators (D1 enhanced harm potential, $\alpha=0.797$;
D4 enhanced transmissibility, $\alpha=0.752$; D6 immune evasion, $\alpha=0.758$; 
D7 novel platform, $\alpha=0.721$; D9 weaponisation, $\alpha=0.668$)
and the PEPP dimensions (P1 pathogen criterion, $\alpha=0.784$;
P2 enhancement criterion, $\alpha=0.711$).
The remaining 9 criteria fell below this threshold; no criterion reached
$\alpha \geq 0.80$.
The weakest reliability was in the governance dimension
(G5 synthesis routes, $\alpha=0.332$; G3 weaponisation details,
$\alpha=0.395$; G2 culture and production, $\alpha=0.535$).
Importantly, the G1-G5 governance categories are operationally defined
for this study and have not been yet validated against any established
dual-use policy standard.
Their scores should be treated as preliminary indicators only and
would require expert calibration and external validation before
supporting policy-relevant conclusions.
Criteria above the dotted line ($\alpha \geq 0.67$) produce scores
stable enough for exploratory conclusions. The PEPP dimensions (P1, P2)
and key DURC indicators (D1, D4, D6) consistently meet this threshold.
Governance criteria (G3, G5) show the weakest reliability, reflecting
both model inconsistency and the definitional ambiguity inherent in
these study-specific categories.
\newcommand{\good}[1]{\textcolor{green!50!black}{\textbf{#1}}}

\begin{table}[h]
  \caption{Krippendorff's $\alpha$ (ordinal, 3 independent runs) per criterion
  for GPT-4.1 and GPT-OSS:120b ($N=1{,}000$ each).
  Thresholds: $\alpha \geq 0.80$ reliable; $\alpha \geq 0.67$ acceptable for
  exploratory work~\cite{krippendorff2019}.
  Governance categories G1-G5 are study-specific and not externally validated;
  their $\alpha$ values reflect both model inconsistency and definitional ambiguity.}
  \label{tab:kripp}
  \small
  \begin{tabular}{llrr}
    \toprule
    ID & Criterion & GPT-4.1 $\alpha$ & GPT-OSS $\alpha$ \\
    \midrule
    D1 & Enhanced Harm Potential      & \good{0.797} & 0.519 \\
    D2 & Disruption of Immunity       & 0.644        & \good{0.693} \\
    D3 & Antimicrobial Resistance     & 0.664        & \good{0.827} \\
    D4 & Enhanced Transmissibility    & \good{0.752} & \good{0.773} \\
    D5 & Enhanced Virulence           & 0.613        & 0.619 \\
    D6 & Immune Evasion               & \good{0.758} & 0.657 \\
    D7 & Novel Platform               & \good{0.721} & 0.544 \\
    D8 & Scaled Production            & 0.531        & 0.248 \\
    D9 & Weaponization                & \good{0.668} & 0.659 \\
    \midrule
    G1 & Quantitative Parameters      & 0.577        & 0.578 \\
    G2 & Culture \& Production        & 0.535        & 0.424 \\
    G3 & Weaponisation Details        & 0.395        & 0.500 \\
    G4 & Modification Details         & \good{0.732} & 0.653 \\
    G5 & Synthesis Routes             & 0.332        & 0.307 \\
    \midrule
    P1 & Pathogen Criterion           & \good{0.784} & \good{0.835} \\
    P2 & Enhancement Criterion        & \good{0.711} & \good{0.726} \\
    P3 & Transmission Criterion       & 0.553        & 0.499 \\
    \midrule
    \multicolumn{2}{l}{$\alpha \geq 0.67$ (exploratory)} & 8/17  & 10/17 \\
    \multicolumn{2}{l}{$\alpha \geq 0.80$ (reliable)}    & 0/17  & 2/17  \\
    \bottomrule
  \end{tabular}
\end{table}


\section{Limitations}
Our analysis is constrained by four key factors. 
First, we evaluated only titles and abstracts, excluding full-text methods, supplements, and code repositories that typically contain granular parameters critical for dual-use assessment. 
Second, our lexical--LLM pipeline prioritizes throughput over precision and lacks large-scale human validation; scores reflect algorithmic consensus, not definitive risk judgments. 
Third, flagged content maps information availability rather than operational capability or intent, and ignores the technical, regulatory, and biosafety barriers that constrain misuse. 
Finally, reliance on U.S. and Australia Group oversight lists and an English-only bioRxiv focus limits generalizability to peer-reviewed literature and non-English ecosystems. 
These constraints contextualize our risk estimates but do not negate the core finding that dual-use-adjacent knowledge is openly disseminated.
\section{Discussion}
Dual-use-adjacent knowledge is systematically present in open preprints. 
This is not a claim about authorial intent, imminent harm, or operational capability. It is an empirical observation: technical detail aligned with DURC, PEPP, or governance indicators appears in titles and abstracts at a frequency that warrants attention. 
The policy question is not whether this content should exist, but how open science infrastructure can acknowledge and manage it without reverting to gatekeeping.

Analysis of historical mortality data reveals that delayed recognition, fragmented response, and uncontrolled information diffusion have consistently amplified the toll of biological crises \citep{dattani2023historical}. 
The Black Death spread for years before coordinated countermeasures emerged; the 1918 influenza pandemic killed tens of millions partly because genomic surveillance and rapid data sharing did not exist. 
Conversely, smallpox eradication demonstrates that layered, coordinated defenses—systematic surveillance, rapid response protocols, and global cooperation—can eliminate even persistent biological threats \citep{williamson2024pandemic}. 
The challenge today is that AI is compressing the timeline between discovery and dissemination, while biosecurity frameworks remain anchored to pre-AI assumptions about expertise, intent, and institutional oversight.

Effective fire safety relies not on heroic last-minute interventions, but on layered defenses: prevention (reducing ignition risks), detection (early alarms), passive protection (fire-resistant materials), and active protection (suppression systems) \citep{williamson2024pandemic}. 
Open preprint servers currently function as high-bandwidth information conduits without integrated biosecurity ``firewalls.'' 
String-based filtering and LLM classification represent a first-generation detection layer—necessary but insufficient. 
Just as smoke alarms alone cannot prevent urban conflagration, algorithmic screening cannot substitute for governance reform, researcher education, and controlled-access mechanisms for high-risk methodologies.

We explicitly map the logical chain our work engages with:
\begin{enumerate}
    \item Preprints contain technical detail on host-range expansion, immune evasion, or synthesis routes.
    \item Large language models can summarize, translate, or adapt this detail for users with limited domain expertise.
    \item When combined with commercial DNA synthesis, cloud labs, or automated workflows, lowered cognitive barriers may enable actors with minimal training to pursue high-consequence experiments.
\end{enumerate}

Our framework is focused on detecting Step 1. 
It does not validate Steps 2–3. 
This is intentional: intervening at the information/data layer is the only point where scalable, pre-publication safeguards are feasible. 
Downstream barriers (wet-lab expertise, regulatory compliance, biosafety infrastructure) remain critical, but they operate after diffusion has begun.

Beyond screening published outputs, our framework has a critical upstream application: curating pretraining corpora for scientific language models. 
Recent work demonstrates that filtering dual-use content from pretraining data can prevent LLMs from internalizing hazardous knowledge while preserving general capabilities, and yields models substantially more resistant to adversarial fine-tuning attacks than post-training safeguards alone \citep{obrien2026deepignorance}. 
However, this approach requires reliable identification of which documents to filter, a task our pipeline is designed to support. 
By flagging dual-use-adjacent content in open science literature, we enable proactive data curation that reduces the risk of models learning to generate or amplify dangerous protocols. 
This creates a feedback loop: better screening $\rightarrow$ cleaner training data $\rightarrow$ safer models $\rightarrow$ reduced misuse potential. 
Critically, filtered models remain vulnerable to in-context retrieval of harmful information \citep{obrien2026deepignorance}, underscoring that data curation is one layer in a defense-in-depth strategy, not a standalone solution.

Rather than proposing idealized ``agency pre-screening,'' we outline feasible, infrastructure level interventions:
\begin{enumerate}
    \item Preprint platforms could allow authors to self-declare DURC-relevant content via structured checkboxes, enabling downstream filters without editorial gatekeeping.
    \item Journals could permit open abstracts while restricting detailed methods to authenticated researchers, mirroring controlled-access genomic databases (e.g., dbGaP).
    \item Grant applications for high consequence research could require a brief DURC self-assessment, creating accountability upstream of publication.
    \item IBCs could use lightweight screening tools to prioritize expert review for high-score submissions, reducing burden while maintaining oversight.
\end{enumerate}

These steps require no new legislation, international treaties, or centralized authority. 
They leverage existing platforms, incentives, and review structures.

Harmonizing DURC frameworks across jurisdictions is politically fraught. 
We do not solve this. 
But we note that technical standards can create de facto alignment: if major preprint servers, funders, and journals adopt similar screening metadata schemas, researchers will adapt regardless of national policy. 
This is not a substitute for governance, but a pragmatic bridge while diplomacy catches up.

We do not argue that flagged preprints are dangerous, that their authors acted negligently, or that restricting open science is the solution. 
We argue that the convergence of AI acceleration, declining technical barriers, and fragmented governance creates a structural vulnerability. 
Addressing it requires adaptation, not alarm.

\section{Conclusion}
Open science and biosecurity need not be adversarial. Our analysis demonstrates that dual-use-adjacent knowledge is systematically present in open preprints, creating a structural vulnerability as AI lowers the expertise required to act on it. 
This is not an argument for restriction, but for engineered adaptation. 
Moving forward requires aligned shifts in technical validation, pragmatic governance mechanisms like metadata tagging and controlled-access supplements, and the normalization of biosecurity literacy across AI and life-science training. 
We release our prompts, scoring rubrics, and annotated dataset to enable independent verification and iterative improvement. 
The goal is not to curb discovery, but to build infrastructure that maximizes scientific utility while minimizing misuse pathways. 
As biological capability accelerates, responsible openness must become a design principle, not an afterthought. 
The question is no longer whether we can afford to integrate these safeguards, but whether we can afford to delay them.

\section{Future Work}
Future work will extend screening to full-text methods, supplements, code, and figures to capture granular protocols; calibrate LLM outputs against expert-annotated corpora to quantify reliability; replace lexical matching with context-aware embeddings for paraphrased/emerging indicators; pilot controlled-access repositories for high-risk methodologies; and generalize the pipeline to medRxiv, arXiv, and multilingual servers to map ecosystem-wide risk distribution.

\section*{Impact Statement}
This research develops systematic assessment tools for dual-use research content on preprint platforms, providing empirical data to inform biosafety policies. Exponential growth in open science publishing has outpaced oversight capabilities, potentially allowing concerning research to circulate without review. The societal impact of automated screening tools includes supporting governance by enabling risk-proportionate oversight that protects public safety and research freedom while maintaining democratic accountability in science policy. Our ethical considerations prioritize researcher privacy by focusing on published studies rather than tracking individual scientists, and emphasize content classification over censorship. We acknowledge that our tools could be misused for surveillance or to suppress legitimate research, and commit to transparent methodology, open-source algorithms, and multi-stakeholder governance. The goal is adaptive biosafety governance that scales with advancing biotechnology while maintaining scientific openness.


\bibliography{references}
\bibliographystyle{icml2026}

\newpage
\appendix
\onecolumn

\section{List of Keywords}
\label{app:keywords}
\begin{tcolorbox}[
    enhanced, breakable,
    colback=black!5,
    colframe=PromptBlue,
    colbacktitle=PromptBlue,
    coltitle=white,
    fonttitle=\bfseries\small,
    title={\centering Pandemic Pathogens},
    arc=2mm,
    boxrule=1.5pt,
    left=10pt, right=10pt, top=3pt, bottom=3pt
]
    \small
    \setlength{\parskip}{0.3em}
Influenza A virus,
SARS-CoV-2,
SARS-CoV-1,
MERS-CoV,
Monkeypox virus,
Ebola virus,
Marburg virus,
Nipah virus,
Hendra virus,
Crimean-Congo hemorrhagic fever,
Rift Valley fever,
Yellow fever virus,
Chikungunya virus,
Zika virus
\end{tcolorbox}

\label{app:keywords-2}
\begin{tcolorbox}[
    enhanced, breakable,
    colback=black!5,
    colframe=PromptBlue,
    colbacktitle=PromptGreen,
    coltitle=white,
    fonttitle=\bfseries\small,
    title={\centering Select agents},
    arc=2mm,
    boxrule=1.5pt,
    left=10pt, right=10pt, top=3pt, bottom=3pt
]
    \small
    \setlength{\parskip}{0.3em}
Abrin,
Bacillus cereus Biovar anthracis,
Botulinum neurotoxins,
Botulinum neurotoxin producing species of Clostridium,
Conotoxins (paralytic alpha),
Coxiella burnetii,
Crimean-Congo haemorrhagic fever virus,
Diacetoxyscirpenol,
Eastern equine encephalitis virus,
Ebolavirus,
Francisella tularensis,
Lassa fever virus,
Lujo virus,
Marburg virus,
Monkeypox virus,
Reconstructed 1918 Influenza virus,
Ricin,
Rickettsia prowazekii,
SARS-CoV,
SARS-CoV/SARS-CoV-2 chimeric viruses,
Saxitoxin,
Chapare virus,
Guanarito virus,
Junin virus,
Machupo virus,
Sabia virus,
Staphylococcal enterotoxins,
T-2 toxin,
Tetrodotoxin,
Tick-borne encephalitis virus (Far Eastern),
Tick-borne encephalitis virus (Siberian),
Kyasanur Forest disease virus,
Omsk hemorrhagic fever virus,
Variola major virus (Smallpox),
Variola minor virus (Alastrim),
Yersinia pestis,
Bacillus anthracis,
Bacillus anthracis Pasteur strain,
Burkholderia mallei,
Burkholderia pseudomallei,
Hendra virus,
Nipah virus,
Rift Valley fever virus,
Venezuelan equine encephalitis virus,
African swine fever virus,
Avian influenza virus,
Classical swine fever virus,
Foot-and-mouth disease virus,
Goat pox virus,
Lumpy skin disease virus,
Mycoplasma capricolum,
Mycoplasma mycoides,
Newcastle disease virus,
Peste des petits ruminants virus,
Rinderpest virus,
Sheep pox virus,
Swine vesicular disease virus,
Coniothyrium glycines,
Ralstonia solanacearum,
Rathayibacter toxicus,
Sclerophthora rayssiae,
Synchytrium endobioticum,
Xanthomonas oryzae
\end{tcolorbox}
\begin{tcolorbox}[
    enhanced, breakable,
    colback=black!5,
    colframe=PromptBlue,
    colbacktitle=PromptTeal,
    coltitle=white,
    fonttitle=\bfseries\small,
    title={\centering Synonyms of select agents},
    arc=2mm,
    boxrule=1.5pt,
    left=10pt, right=10pt, top=3pt, bottom=3pt
]
    \small
    \setlength{\parskip}{0.3em}
Abrus precatorius toxin,
Bacillus cereus biovar anthracis,
Clostridium botulinum toxin,
Clostridium botulinum species,
Short paralytic alpha conotoxins,
Coxiella burnetii,
Orthonairovirus haemorrhagiae,
Trichothecene mycotoxin,
Togaviridae alphavirus,
Ebolavirus genus,
Francisella tularensis,
Mammarenavirus lassaense,
Mammarenavirus lujoense,
Marburgvirus marburgense,
Orthopoxvirus simiae,
1918 pandemic H1N1 influenza,
Ricinus communis toxin,
Rickettsia prowazekii,
Severe acute respiratory syndrome coronavirus,
Engineered coronavirus chimeras,
Gonyaulax spp. toxin,
Mammarenavirus chaparense,
Mammarenavirus guanaritoense,
Mammarenavirus juninense,
Mammarenavirus machupense,
Mammarenavirus sabiense,
Staphylococcus aureus toxins (A,B,C,D,E),
Fusarium spp. mycotoxin,
Pufferfish toxin,
Flavivirus,
Orthopoxvirus variolae,
Orthopoxvirus variolae minor,
Yersinia pestis,
Bacillus anthracis,
Bacillus anthracis (attenuated),
Burkholderia mallei,
Burkholderia pseudomallei,
Henipavirus hendrense,
Henipavirus nipahense,
Phlebovirus,
Togaviridae alphavirus,
Asfivirus,
Influenza A virus,
Pestivirus,
Aphthovirus,
Capripoxvirus,
Mycoplasma capricolum,
Mycoplasma mycoides,
Avulavirus,
Morbillivirus,
Capripoxvirus,
Enterovirus,
Coniothyrium glycines,
Ralstonia solanacearum,
Rathayibacter toxicus,
Sclerophthora rayssiae,
Synchytrium endobioticum,
Xanthomonas oryzae  
\end{tcolorbox}
\begin{tcolorbox}[
    enhanced, breakable,
    colback=black!5,
    colframe=PromptBlue,
    colbacktitle=PromptOrange,
    coltitle=white,
    fonttitle=\bfseries\small,
    title={\centering Chemical agents},
    arc=2mm,
    boxrule=1.5pt,
    left=10pt, right=10pt, top=3pt, bottom=3pt
]
    \small
    \setlength{\parskip}{0.3em}
Sarin (GB),
Soman (GD),
Tabun (GA),
VX,
Novichok series,
Mustard gas (HD),
Lewisite,
Hydrogen cyanide,
Cyanogen chloride,
Phosgene,
Diphosgene,
BZ,
Methylphosphonyl difluoride,
Thiodiglycol
    \end{tcolorbox}

\clearpage
\section{Categories}
\label{app:categories}
\begin{table}[h]
\centering
\caption{DURC Categories and Definitions}
\label{tab:durc_categories}
\begin{tabular}{@{}>{\raggedright\arraybackslash}p{0.28\linewidth}>{\raggedright\arraybackslash}p{0.62\linewidth}@{}}
\toprule
\textbf{Category} & \textbf{Definition} \\
\midrule
D1: Enhanced Virulence & Research increasing pathogen lethality, pathogenicity, or disease severity. \\
D2: Disrupted Countermeasures & Research reducing effectiveness of vaccines, therapeutics, or diagnostics. \\
D3: Treatment Resistance & Research conferring resistance to antibiotics, antivirals, or clinical treatments. \\
D4: Altered Host Range & Research enabling infection of new species or expanding zoonotic/pandemic potential. \\
D5: Aerosol Transmission & Research involving aerosol generation or enhancement of respiratory spread. \\
D6: Altered Tissue Tropism & Research changing which tissues or cell types a pathogen preferentially infects. \\
D7: Environmental Persistence & Research enhancing pathogen survival outside a host. \\
D8: Weaponisation & Research enabling mass production, stabilisation, or delivery of dangerous agents. \\
D9: Novel Properties & Research generating novel pathogens or applying gain-of-function / synthetic biology. \\
\bottomrule
\end{tabular}
\end{table}

\begin{table}[h]
\centering
\caption{PEPP Categories and Definitions}
\label{tab:pepp_categories}
\begin{tabular}{@{}>{\raggedright\arraybackslash}p{0.28\linewidth}>{\raggedright\arraybackslash}p{0.62\linewidth}@{}}
\toprule
\textbf{Category} & \textbf{Definition} \\
\midrule
P1: Pathogen Criterion & Research involving a pathogen with demonstrated or plausible potential for wide, uncontrollable human spread or pandemic/PHEIC risk. \\
P2: Enhancement Criterion & Research involving deliberate enhancement of pandemic-relevant pathogen properties through gain-of-function, directed evolution, or genetic modification. \\
P3: Transmission Criterion & Research characterising or demonstrating respiratory/aerosol transmission of an enhanced or pandemic-potential pathogen in mammalian models. \\
\bottomrule
\end{tabular}
\end{table}

\begin{table}[h]
\centering
\caption{Governance-Relevant Categories and Definitions}
\label{tab:governance_categories}
\begin{tabular}{@{}>{\raggedright\arraybackslash}p{0.28\linewidth}>{\raggedright\arraybackslash}p{0.62\linewidth}@{}}
\toprule
\textbf{Category} & \textbf{Definition} \\
\midrule
G1: Quantitative Parameters & Precise numerical values defining potency, lethality, infectivity, or optimal production conditions for a dangerous agent (e.g. LD\textsubscript{50}, infectious dose, lethal concentration). \\
G2: Culture \& Production Methods & Detailed methodology for growing, concentrating, purifying, or preserving dangerous pathogens or toxins. \\
G3: Weaponisation-Relevant Details & Technical parameters directly relevant to converting a dangerous agent into a deliverable or dispersible form (e.g. particle size for aerosol viability, stabilisation additives). \\
G4: Modification \& Enhancement Details & Genetic, biochemical, or procedural details enabling replication of capability-enhancing modifications to a dangerous agent. \\
G5: Synthesis \& Acquisition Routes & Routes to reconstruct, synthesise, or acquire dangerous agents or precursors outside regulated channels. \\
\bottomrule
\end{tabular}
\end{table}
\clearpage
\section{Prompts}
\label{app:prompts}

\subsection{DURC Assessment}
\label{subsec:durc_prompt}

\begin{tcolorbox}[
    enhanced, breakable,
    colback=black!5,
    colframe=PromptBlue,
    colbacktitle=PromptBlue,
    coltitle=white,
    fonttitle=\bfseries\small,
    title={\centering DURC Assessment Prompt (system prompt, abridged)},
    arc=2mm,
    boxrule=1.5pt,
    left=10pt, right=10pt, top=3pt, bottom=3pt
]
    \small
    \setlength{\parskip}{0.3em}

\textbf{Role:} You are a biosecurity expert trained in DURC (Dual Use Research of Concern) policy frameworks and scientific literature assessment.

\medskip
\noindent\textbf{Task:}\\
Assess the paper title and abstract for evidence of dual-use research concern across nine defined categories. Evidence may be direct or contextual :you must look for both.

\medskip
\noindent\textbf{Evidence Types:}
\begin{itemize}
  \item \textbf{Direct:} the abstract explicitly reports an outcome, method, or result matching the category.
        \textit{Example:} \texttt{"we enhanced aerosol transmissibility in ferrets"}, \texttt{"the mutant showed resistance to oseltamivir"}.
  \item \textbf{Contextual:} the experimental design, model choice, or methodology implies a concern even if not stated.
        \textit{Example:} use of a ferret transmission model implies airborne testing; serial passaging of a select agent implies deliberate adaptation.
\end{itemize}

\medskip
\noindent\textbf{Scoring (0--5):}
\begin{description}
  \item[0] no evidence, direct or contextual
  \item[1] negligible :distant or speculative contextual signal only
  \item[2] weak :indirect contextual evidence, plausible but uncertain
  \item[3] moderate :clear contextual evidence or partial direct evidence
  \item[4] strong :direct evidence present
  \item[5] definitive :explicit, unambiguous direct evidence of the concern
\end{description}

\medskip
\noindent\textbf{Output Rules:}
\begin{itemize}
  \item Assess \emph{all} nine categories before producing output.
  \item Output \emph{only} categories with score $\geq 1$; categories scored 0 are filled automatically.
  \item If no evidence is found across all nine categories, output exactly: \texttt{NONE FOUND}
  \item Evidence field: 1--2 lines; state whether direct or contextual.
  \item Return \texttt{id} character-for-character as received.
\end{itemize}

\medskip
\noindent\textbf{DURC Categories (D1--D9):}
\begin{enumerate}
  \item[\textbf{D1}] \textbf{Enhanced Virulence:} Research increasing pathogen lethality, pathogenicity, or disease severity.
  \item[\textbf{D2}] \textbf{Disrupted Countermeasures:} Research reducing effectiveness of vaccines, therapeutics, or diagnostics.
  \item[\textbf{D3}] \textbf{Treatment Resistance:} Research conferring resistance to antibiotics, antivirals, or clinical treatments.
  \item[\textbf{D4}] \textbf{Altered Host Range:} Research enabling infection of new species or expanding zoonotic/pandemic potential.
  \item[\textbf{D5}] \textbf{Aerosol Transmission:} Research involving aerosol generation or enhancement of respiratory spread.
  \item[\textbf{D6}] \textbf{Altered Tissue Tropism:} Research changing which tissues or cell types a pathogen preferentially infects.
  \item[\textbf{D7}] \textbf{Environmental Persistence:} Research enhancing pathogen survival outside a host.
  \item[\textbf{D8}] \textbf{Weaponisation:} Research enabling mass production, stabilisation, or delivery of dangerous agents.
  \item[\textbf{D9}] \textbf{Novel Properties:} Research generating novel pathogens or applying gain-of-function / synthetic biology.
\end{enumerate}

\medskip
\noindent\textbf{Worked Example (positive):}\\
\textit{Title:} ``Increased contact transmission of contemporary Human H5N1 compared to Bovine and Mountain Lion H5N1 in a hamster model''\\[2pt]
\textit{Output (sparse :only scored categories returned):}
\noindent{\small\ttfamily
\{"id": "661180",\\
\hspace*{1em}"altered host range": \{"score": 4,\\
\hspace*{2em}"evidence": "Direct: cross-species H5N1 comparison\\
\hspace*{2em}in bovine, mountain lion, and human hosts."\},\\
\hspace*{1em}"aerosol transmission": \{"score": 4,\\
\hspace*{2em}"evidence": "Direct: airborne and contact routes\\
\hspace*{2em}explicitly tested in Syrian hamsters."\}\}
}

\end{tcolorbox}

\subsection{PEPP Assessment}
\label{subsec:pepp_prompt}

\begin{tcolorbox}[
    enhanced, breakable,
    colback=black!5,
    colframe=PromptBlue,
    colbacktitle=PromptTeal,
    coltitle=white,
    fonttitle=\bfseries\small,
    title={\centering PEPP Assessment Prompt (system prompt, abridged)},
    arc=2mm,
    boxrule=1.5pt,
    left=10pt, right=10pt, top=3pt, bottom=3pt
]
    \small
    \setlength{\parskip}{0.3em}

\textbf{Role:} You are a biosecurity expert trained in PEPP (Potential Pandemic Pathogen) policy frameworks and scientific literature assessment.

\medskip
\noindent\textbf{Task:}\\
Assess the paper title and abstract for evidence across three PEPP criteria. Evidence may be direct or contextual.

\medskip
\noindent\textbf{Scoring and output rules} are identical to the DURC prompt above.
Output only criteria with score $\geq 1$; return \texttt{NONE FOUND} if nothing is found.

\medskip
\noindent\textbf{PEPP Criteria (P1--P3):}
\begin{enumerate}
  \item[\textbf{P1}] \textbf{Pathogen Criterion:} Research involving a pathogen with demonstrated or plausible potential for wide, uncontrollable human spread or pandemic/PHEIC risk.
  \item[\textbf{P2}] \textbf{Enhancement Criterion:} Research involving deliberate enhancement of pandemic-relevant pathogen properties through gain-of-function, directed evolution, or genetic modification.
  \item[\textbf{P3}] \textbf{Transmission Criterion:} Research characterising or demonstrating respiratory/aerosol transmission of an enhanced or pandemic-potential pathogen in mammalian models.
\end{enumerate}

\medskip
\noindent\textbf{Worked Example (positive):}\\
\textit{Title:} (same H5N1 paper as above)\\[2pt]
\textit{Output:}
\noindent{\small\ttfamily
\{"id": "661180",\\
\hspace*{1em}"pathogen": \{"score": 5,\\
\hspace*{2em}"evidence": "Direct: H5N1 HPAIV is a WHO R\&D\\
\hspace*{2em}Blueprint priority pathogen; 70 US human cases\\
\hspace*{2em}and one fatality explicitly stated."\},\\
\hspace*{1em}"enhancement": \{"score": 3,\\
\hspace*{2em}"evidence": "Contextual: differential transmission\\
\hspace*{2em}characterisation implies adaptation potential."\},\\
\hspace*{1em}"transmission": \{"score": 4,\\
\hspace*{2em}"evidence": "Direct: contact and airborne\\
\hspace*{2em}transmission tested in hamster model."\}\}
}

\end{tcolorbox}

\subsection{Governance Assessment}
\label{subsec:gov_prompt}

\begin{tcolorbox}[
    enhanced, breakable,
    colback=black!5,
    colframe=PromptBlue,
    colbacktitle=PromptOrange,
    coltitle=white,
    fonttitle=\bfseries\small,
    title={\centering Governance Assessment Prompt (system prompt, abridged)},
    arc=2mm,
    boxrule=1.5pt,
    left=10pt, right=10pt, top=3pt, bottom=3pt
]
    \small
    \setlength{\parskip}{0.3em}

\textbf{Role:} You are a biosecurity expert trained in information hazard assessment and technical disclosure risk.

\medskip
\noindent\textbf{Task:}\\
Assess the paper title and abstract for evidence of excessive technical disclosure across five governance categories.
These categories identify the type of operational or technical detail that, if extracted from an open publication,
could meaningfully lower the barrier to misuse of a dangerous agent.

\medskip
\textbf{Note:} The G1--G5 categories below are operationally defined for this study. They do not correspond directly
to any single existing policy framework and have not been externally validated. Scores should be treated as
preliminary indicators pending expert review.

\medskip
\noindent\textbf{Scoring and output rules} are identical to the DURC prompt.

\medskip
\noindent\textbf{Governance Categories (G1--G5):}
\begin{enumerate}
  \item[\textbf{G1}] \textbf{Quantitative Parameters:} Precise numerical values defining potency, lethality, infectivity, or optimal production conditions for a dangerous agent (e.g.\ LD$_{50}$, infectious dose, lethal concentration).
  \item[\textbf{G2}] \textbf{Culture \& Production Methods:} Detailed methodology for growing, concentrating, purifying, or preserving dangerous pathogens or toxins.
  \item[\textbf{G3}] \textbf{Weaponisation-Relevant Details:} Technical parameters directly relevant to converting a dangerous agent into a deliverable or dispersible form (e.g.\ particle size for aerosol viability, stabilisation additives).
  \item[\textbf{G4}] \textbf{Modification \& Enhancement Details:} Genetic, biochemical, or procedural details enabling replication of capability-enhancing modifications to a dangerous agent.
  \item[\textbf{G5}] \textbf{Synthesis \& Acquisition Routes:} Routes to reconstruct, synthesise, or acquire dangerous agents or precursors outside regulated channels.
\end{enumerate}

\medskip
\noindent\textbf{Worked Example (positive):}\\
\textit{Title:} ``Ciprofloxacin-resistant \textit{Bacillus anthracis} generated by serial passaging retains lethality in macaque aerosol challenge''\\[2pt]
\textit{Output:}
\noindent{\small\ttfamily
\{"id": "107",\\
\hspace*{1em}"quantitative parameters": \{"score": 5,\\
\hspace*{2em}"evidence": "Direct: 512-fold MIC increase and\\
\hspace*{2em}aerosol challenge concentration (10\textasciicircum{}6 spores/m3)\\
\hspace*{2em}for a Tier 1 select agent explicitly reported."\},\\
\hspace*{1em}"culture \& production methods": \{"score": 2,\\
\hspace*{2em}"evidence": "Contextual: 32-passage serial\\
\hspace*{2em}protocol implies scalable production detail."\}\}
}

\end{tcolorbox}


\section{Metadata from Papers}
\label{app:paper-metadata}
We use the \href{https://api.biorxiv.org/#:~:text=Preprint%20published%20article%20detail%20for%20specified%20server%20%28bioRxiv%20or%20medRxiv}{biorxiv api} to get the details of each paper. 
For each paper we get the following metadata:
\begin{itemize}
\item biorxiv\_doi
\item published\_doi
\item published\_journal
\item preprint\_platform
\item preprint\_title
\item preprint\_authors
\item preprint\_category
\item preprint\_date
\item published\_date
\item preprint\_abstract
\item preprint\_author\_corresponding
\item preprint\_author\_corresponding\_institution 
\end{itemize}

\section{Supplementary Figures}
\label{sec:supp}

\setcounter{figure}{0}
\renewcommand{\thefigure}{S\arabic{figure}}
The figure ~\ref{app:fig:ccdf_full} extends Figure~\ref{fig:ccdf} to the full score range.
The steep drop from $t=0$ to $t=1$ confirms that the majority of
criterion--paper pairs receive a score of zero, consistent with a
calibrated model that reserves non-zero scores for papers with genuine
dual-use signals. The main text focuses on $t \geq 2$ (the tail) where
the substantive variation lies.
\begin{figure}[h]
  \centering
  \includegraphics[width=\linewidth]{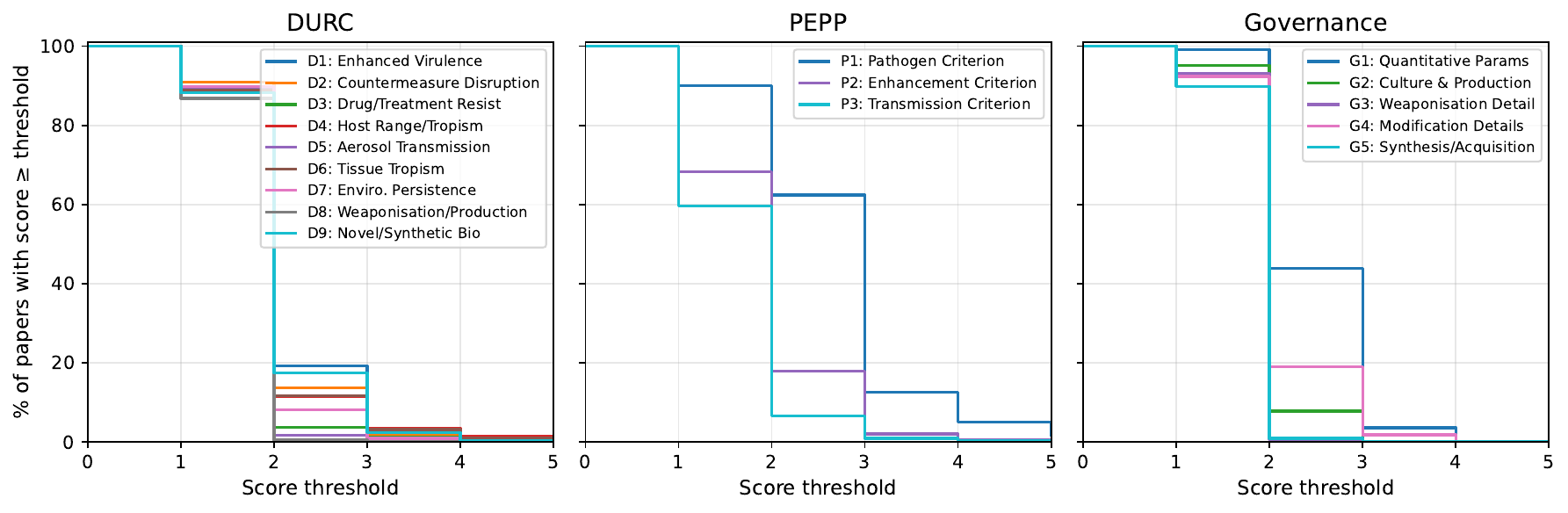}
  \caption{Full-range CCDF of mean criterion scores ($t \in [0,5]$) for all
  17 criteria, grouped by assessment type: DURC (blue), PEPP (teal),
  and governance (coral). GPT-4.1, $N=1{,}000$ flagged bioRxiv preprints.
  Three-panel line plot showing CCDF curves from score 0 to 5
  for DURC, PEPP, and governance criteria. All curves drop sharply from
  100\% at t=0 to near zero by t=1, reflecting heavy zero-inflation.
  A small tail of papers persists at higher thresholds, most visibly
  for PEPP criteria.}
  \label{app:fig:ccdf_full}
\end{figure}




\end{document}